\documentclass[12pt]{article}
\usepackage{epsfig}
\begin{document}
\vspace*{-.6in} \thispagestyle{empty}
\begin{flushright}
SUSX-TH/01-038

IPPP/01/36

DCPT/01/72
\end{flushright}
\baselineskip = 20pt

\vspace{.5in} {\Large
\begin{center}
{\bf CP Violation and Dilaton Stabilization in Heterotic String Models}

\end{center}}

\vspace{.5in}

\begin{center}
Shaaban Khalil$^{a,b}$,
  Oleg Lebedev$^{c}$ and Stephen
Morris$^{c}$ \\

\vspace{.5in}

a: \emph{IPPP, University of Durham, South Rd., Durham DH1 3LE, U.K. }
\\

b: \emph{Ain Shams University, Faculty of Science, Cairo, 11566, Egypt }\\

c: \emph{Centre for Theoretical Physics, University of Sussex,\\
Falmer, Brighton, BN1 9QJ, U.K.}
\end{center}

\vspace{.5in}

\begin{abstract}
We study the possibility of spontaneous CP violation  in string
models with the dilaton field stabilized at a phenomenologically
acceptable  value. We consider three mechanisms to stabilize the dilaton:
multiple gaugino condensates, a nonperturbative K\"{a}hler potential,
and a superpotential based on S-duality, and analyze consequent CP
phases in the soft SUSY breaking terms.
Due to non-universality forced upon the theory by requiring a non-trivial
CKM phase, the EDM problem becomes more severe. Even if  there are no complex
phases  in the VEVs of the SUSY breaking fields, the electric
dipole moments are overproduced by orders of magnitude.
We also address the question of modular invariance of the physical CP phases.
\end{abstract}

\noindent

\newpage


\section{Introduction}
At present, string theory and its extension, M-theory, provide
the most promising schemes for the unification of the fundamental
forces of nature. But before we can assign it the status of a
Theory of Everything, we need to establish that the phenomenology
of the Standard Model (SM) can be produced by some reasonable
scenario. One of the outstanding problems is the origin of CP
violation. In string theory CP is a gauge symmetry and can only be
broken spontaneously \cite{Dine:1992ya} by a vacuum expectation
value (VEV) of some SM-singlet field. There is a variety of good
candidates to do this job. These include  the dilaton, $S$, which
is related to the gauge coupling via
\begin{equation}
\label{dilaton}
{\rm  Re}S \simeq \frac{1}{g^{2}_{\rm {string}}},
\end{equation}\
and the moduli which parameterize the size and shape of the
compact dimensions, $T_i$ and  $U_i$ (i=1,2,3). All of these
generically attain complex VEVs and  induce spontaneous breakdown
of the CP symmetry \cite{Strominger:1985it}. One must, however,
make sure that this produces the right amount of CP violation. For
instance, Im$S$ of order one induces very large electric dipole
moments (EDMs) through a contribution to the QCD $\bar \theta$
parameter. On the other hand, it does not induce the
Cabibbo-Kobayashi-Maskawa (CKM) phase. Therefore the dilaton alone
can hardly do a good job. The  moduli fields can, at least in
principle, induce the CKM phase without violating the EDM bounds.
In this paper we will consider the moduli fields $T_i$ as the
source of observable CP violation and study to what extent such a
scenario is phenomenologically viable\footnote{Throughout this
paper we shall assume that all moduli $T_i$ have the same value,
$T$.}.

An important property of weakly coupled heterotic  orbifold
models is
the existence of an $SL(2,\mathbf{Z}) $ symmetry:
with integer $a,b,c,d$ and $ad-bc=1$.
The dilaton field is inert under this transformation at the tree level,
but at the one loop level the cancellation of the modular anomaly
requires it to transform  as (up to a T-independent imaginary shift)
\begin{equation}\label{strans}
  S \longrightarrow S+\frac{3}{4 \pi^2} \delta_{GS} \ln (i c
  T+d)\;,
\end{equation}
where $\delta_{GS}$ is the Green-Schwarz coefficient. In the
weakly coupled heterotic string theory this symmetry is preserved
to all orders in perturbation theory and provides  important
guidance for constructing low energy effective field theories. One
however should keep in mind that this symmetry is broken
spontaneously below the compactification scale and is realized
nonlinearly at the electroweak scale. The modular symmetry has
important implications for CP violation as it sometimes allows us
to eliminate unphysical CP phases.

Complex VEVs of the moduli fields as a source of CP violation have
been considered before, see Ref.\cite{Bailin:1998iz}
and references therein. These models however are not fully
realistic as they do not address the issue of dilaton
stabilization. The observed unification of the gauge coupling
constants at $\alpha({\rm {GUT}})=g^2/4 \pi= 1/25$ implies that
$\mathrm{Re} S$ should take a value of around two. So, a
consistent model must satisfy this requirement, which also has
important implications for supersymmetry breaking. In this paper
we shall examine scenarios which satisfy the following three
requirements:
\\ \  \\
{\bf 1.} The dilaton field is stabilized at Re$S \sim 2$.\\
{\bf 2.} CP is violated.\\
{\bf 3.} The supersymmetry breaking scale is phenomenologically
acceptable.
\\ \ \\
We find that these requirements are quite stringent and leave very
few viable possibilities. A somewhat similar question in the context
of effective Type I models was addressed in Ref.\cite{Abel:2001ur}.
We note that one can also impose an additional phenomenological constraint
that the  flavor changing neutral currents are absent.
However, as we will see, this constraint is often satisfied automatically
in the class of models under consideration.

The paper is organized as follows. In section 2 we review supersymmetry breaking
via gaugino condensation and discuss  properties of the scalar
potential possessing a modular symmetry. In section 3 we present our analysis
of dilaton stabilization via multiple gaugino condensates, non-perturbative
K\"{a}hler potential, and S-duality.
The CKM phase in heterotic models is discussed in section 4.
Section 5 is devoted to the discussion
of the soft terms and modular properties of the CP phases.
In section 6 we analyze various types of EDM contributions encountered in our
models. Finally, the conclusions are presented in section 7.

 All of the models we examine start with a
 single hidden sector gaugino condensate and then modify it
 to produce dilaton stabilization, so now we will take a short
 detour and give an introduction to this simple case.

\section{Gaugino Condensation}

Hidden sector gaugino condensation is one of the most popular
schemes for the breaking of supersymmetry (see \cite{Bailin:1999nk}
for a recent review).  This is realized in
$E_8 \otimes E_8$ heterotic string theory where the condensate
lives in one $E_8$, the other forming the observable sector. The
Veneziano-Yankielowicz superpotential which describes the
condensate is given by \cite{Veneziano:1982ah}:
\begin{equation}\label{YYsup}
W=\frac{1}{4}U \Big(f+\frac{2}{3} \beta \ln U \Big)\;,
\end{equation}
where $\beta$ is the one-loop coefficient of the beta
function, $U=\delta_{ab}
W^a_{\alpha}\epsilon^{\alpha\beta}W^b_{\beta}$ is a chiral
superfield  whose lowest component corresponds to the gaugino
condensate $\langle \lambda\lambda \rangle$, and
\begin{equation}\label{ff}
  f=S+ \left( 4 \beta -\frac{ 3 \delta_{GS}}{2 \pi^2} \right) \ln \eta(T)
\end{equation}
is the gauge kinetic function.

In this paper we shall use a truncated superpotential, found by
replacing $U$ by its value at $\frac{\partial W}{\partial U}=0$.
This approximates its value at the minimum of the scalar potential
\cite{deCarlos:1991gq} and gives:
\begin{equation}\label{trunW}
  W=d \frac{e^{\frac{-3 S}{2\beta}}}{\eta(T)^{6-\frac{9 \delta_{GS}}{4
  \pi^2 \beta}}}
\end{equation}
with $d=-\beta /6e$. The (standard)  K\"{a}hler potential is given
by \cite{Witten:1985xb}:
\begin{equation}\label{K}
  K=-\ln Y-3\ln(T+\overline{T})\;,
\end{equation}
where $Y=S+\overline{S}+\frac{3}{4
\pi^2}\delta_{GS}\ln(T+\overline{T})$. The  scalar potential 
is expressed as
\begin{equation}\label{V}
  V=e^G\left( G_i \left( G^i_j \right)^{-1}G^j-3 \right)\;,
  \end{equation}
where $G=K+\ln(|W|^2)$ and the subscripts (superscripts)
denote differentiation and the sum over repeated indices runs over
the (conjugate) fields in the system. Supersymmetry is broken by
VEVs of the auxiliary fields  ($j=S,T$):
\begin{equation}\label{F}
 F_j =e^{G/2}\left(G^i_j\right)^{-1} G_i\;,
  \end{equation}
String models typically contain hidden matter which can couple to
the condensate. For scalar matter fields, $A$, with a gauge group
$SU(N)$ and ``quarks" with $M~(N+\overline{N})$ representations,
the condensate superpotential is:
\begin{equation}\label{Wmat}
  W(S,T,A)=-N(32 \pi^2 e)^{\frac{M}{N}-1}(\det
  \mathcal{M})^{\frac{1}{N}}\frac{e^{-\frac{8 \pi^2
  S}{N}}}{\eta(T)^{\frac{32 \pi^2 \beta}{ N}-\frac{12
  \delta_{GS}}{N}}}\;,
\end{equation}
 where $\mathcal{M}$ is a matrix containing the
coefficients of the quarks in their trilinear terms in the
superpotential, $\sum_{r, a,b}h_{rab}A_rQ_a
\overline{Q}_b=\mathcal{M}_{ab}Q_a \overline{Q}_b$. We assume a
generic singlet field $A=A_r$, so $\det\mathcal{M}=A^M$.

Since for a realistic case $\langle A^2 \rangle \ll S_R , T_R$,
the scalar potential is dominated by the term proportional to
$\vert \partial W/ \partial A \vert^2$ \cite{deCarlos:1993da}.
So, to a good approximation, the minimum occurs at $\frac{\partial
W}{\partial A}=0$, and we can neglect the
terms containing $A$ in the K\"{a}hler potential. Then we have:
\begin{equation}\label{Wsimp}
  W=\tilde d \frac{e^{\frac{-3 S}{2 \tilde\beta}}}{\eta(T)^{6-\frac{9 \delta_{GS}}{4
\pi^2\tilde\beta}}}
\end{equation}
where $\tilde\beta=\frac{3N-M}{16 \pi^2}$ is the beta function and
$\tilde d=(M/3-N)(32 \pi^2 e)^{\frac{3 (M-N)}{3N-M}}(
M/3)^{\frac{M}{3N-M}}$.

This model does not lead to dilaton stabilization at a reasonable
value \cite{deCarlos:1993da}, in fact at the minimum ${\rm Re}S \rightarrow \infty$.
So we must consider modifications.
We shall study three models, one where corrections are made to the
K\"{a}hler potential, one with an S-dual potential and another
with two gaugino condensates\footnote{Another dilaton stabilization 
mechanism, in the context of type I models, was suggested in \cite{Abel:2001tf}. }.

Before we proceed, let us mention a few useful facts. First, if the scalar potential possesses
an $SL(2,{\rm Z})$ symmetry, the fixed points under the duality group are always
stationary. In practice, these fixed points are often minima. As we will see, if the modulus
field is stabilized at a fixed point, the CKM phase vanishes and often there is no
supersymmetry breaking. So our task will be to pull the minima away from the fixed points.

Second, for $\delta_{GS}=0$ and the standard K\"{a}hler potential
(\ref{K}) , $F_T=0$ at the T-duality fixed points while $F_S=0$ at
the S-duality fixed points. This follows from the fact that $F_T
\propto G_{ T}$ and
\begin{equation}
G_{T} \propto {1\over T+\bar T}+ 2{\eta'(T) \over \eta(T)}\biggl\vert_{\rm f.p.}=0\;,
\label{G}
\end{equation}
keeping in mind that a derivative of a modular invariant function vanishes at the fixed
points.
This, of course, equally applies to the S-dual potentials.

Third, let us mention a useful
property of models with
factorizable effective superpotentials, i.e. $W_{eff}=\Omega(S)/\Lambda(T)$.
In such models, for $\delta_{GS}=0$ an extremum (which is often a minimum)
of the scalar potential occurs at
\begin{equation}
2S_R W_S-W=0\;,
\label{Fs}
\end{equation}
assuming the standard K\"{a}hler potential (\ref{K}) \footnote{This also applies to
$T$ and $W_T$. }.
Consequently, $F_S=0$ at this point since it is proportional to
precisely this combination. In the general case of
$\delta_{GS}\not= 0$, there can be departures from this result.

Finally, in what follows we will  consider generalized superpotentials
consistent with the modular symmetry. That is, we will use the freedom
to multiply the superpotential arising from gaugino condensation by
a modular invariant function $H(T)$\cite{Cvetic:1991qm}:

\begin{equation}\label{H}
  H(T)=\Bigl[j(T)-1728\Bigr]^{\frac{m}{2}}j(T)^{\frac{n}{3}}P\left[j(T)\right]\;,
\end{equation}\

\noindent where $j(T)$ is the absolutely modular invariant
function, $P\left[j(T)\right]$ is some polynomial of $j(T)$, and
$m,n$ are integers. This is the most general modification
consistent with T-duality and absence of singularities inside the
fundamental domain. However, one should keep in mind that explicit
examples where $H(T)$ appears are lacking, so it is possible that
we allow ourselves more freedom than there is in practise.

We will typically set $P\left[j(T)\right]$ to
one, noting that increasing the amount of $j(T)$ in the potential
typically forces the minima of $T$ to the fixed points. In
practice, the modulus field often gets stabilized at $T_{\rm min
}=1$. Then, for $m>0$,  there is no supersymmetry breaking because
$H(1)\bigl\vert_{m>0}=0$, and such models must be discarded.
The same applies to the other fixed points for $n>0$ since
$H(e^{\pm i \pi/6})\bigl\vert_{n>0}=0$.


\section{Dilaton Stabilization}

As we have seen above, the simplest model of gaugino condensation does not lead to a
finite value of the dilaton field. Since the VEV of the dilaton describes the
gauge coupling constants, such a  model is phenomenologically unacceptable.
The simple model above might well be oversimplified and in more involved
models dilaton stabilization can be achieved while retaining the main features
of the single gaugino condensate model.


\begin{figure}[ht]
\begin{center}
\begin{tabular}{c}

\epsfig{file=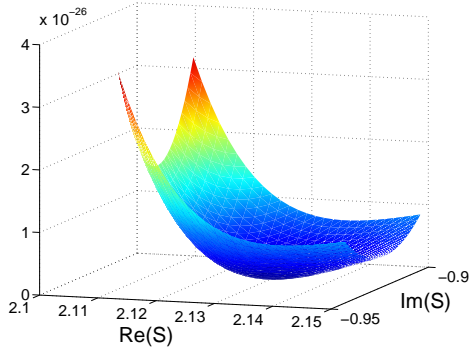, height=5cm}
\end{tabular}
\end{center}

\caption{Racetrack scalar potential with $H$ and m=1, n=0.
$T$ is set to its minimum value, $T_{min}=0.9850e^{0.5471i}$. The minimum in $S$ is
at $S_{min}=2.13-0.92i$.
}
\label{CPfig1}
\end{figure}

\begin{figure}[ht]
\begin{center}
\begin{tabular}{c}

\epsfig{file=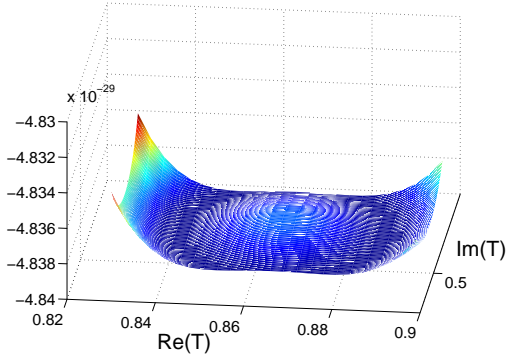, height=5cm}

\end{tabular}
\end{center}

\caption{Racetrack scalar potential with $H$ and m=1, n=0.
$S$ is set to its minimum value, $2.13-0.92i$.The minimum in $T$ is
at $T_{min}=0.9850e^{0.5471i}$.}
\label{CPfig2}
\end{figure}

\subsection{Models With Two Gaugino Condensates (``Racetrack Models'')}

Generally, the hidden sector may contain non-semi-simple gauge groups.
Given the right matter content, it is plausible that gauginos condense
in each of the simple group factors. With a nonzero Im$S$, these condensates
may enter the superpotential with opposite signs thereby leading to dilaton
stabilization.

Let us consider a model with two  gaugino condensates.
Suppose  we have a  gauge group $SU(N_1)\otimes SU(N_2)$
with $M_1(N_1+\overline{N}_1)$ and $M_2(N_2+\overline{N}_2)$ ``quark"
representations. The superpotential is simply the sum of that,
(\ref{Wsimp}), for each of the individual condensates
\cite{deCarlos:1993da}:
\begin{equation}
 W=\tilde d_1 \frac{e^{\frac{-3 S}{2 \tilde\beta_1}}}{\eta(T)^{6-\frac{9 \delta_{GS}}{4
\pi^2\tilde\beta_1}}}+
\tilde d_2 \frac{e^{\frac{-3 S}{2 \tilde\beta_2}}}{\eta(T)^{6-\frac{9 \delta_{GS}}{4
\pi^2\tilde\beta_2}}}\;.
\end{equation}
For  $N_1=6, N_2=7$ and $M_1=2, M_2=8$ we have dilaton
stabilization at $S=2.1-0.92i$ and $T=1.23$ \cite{Barreiro:1998aj}.
In this case the CKM phase is zero and we should consider modifications.
Keeping the same condensing gauge groups and the matter content,
we can  multiply the superpotential by $H(T)$.
The resulting minima are
shown in Table \ref{table1}.

The table shows $S$ and $T$ at the minima along with their auxiliary
fields for various powers $m$ and $n$.
SUSY breaking is given in Planck units and $0$ indicates
SUSY breaking much below the phenomenologically allowed range.
For $\delta_{GS}=0$,
we see that  $S$ is always stabilized  at a reasonable value and
$T$ is complex for $m\geq 1$. This is illustrated in Figs.\ref{CPfig1} and \ref{CPfig2}.
We can see that for $m=1$,$n=0$ the fixed point $e^{i\pi/6}$ is a local maximum.

In all cases $F_S=0$ due to Eq.\ref{Fs}, whereas $F_T$ may be nonzero.
In most cases the modulus is stabilized
on the unit circle and SUSY remains unbroken.
The presence of extrema on the unit circle can be seen from the fact that
that there is always a stationary point at $G_S=G_T=0$ and 
there is a point on the unit circle where $G_T$ vanishes 
(since it is a derivative of a modular invariant function),
whereas $G_S$ is always zero. The minima on the unit circle (if they exist) 
away from the fixed points are typically lower then those at the fixed points
because in the former case SUSY breaking is zero and the potential is
negative while in the latter case the entire potential vanishes.

For $m=n=0$ and
$m= 1,2$, $n=0$ we have viable supersymmetry breaking
($10^2~GeV \leq F_{T} \leq 10^4~GeV$), but only in
the latter case is there  CP violation. However, even in this
case $T$ is close to the fixed point $e^{i\pi/6}$  which results
in a suppressed ($\sim{\cal O}(0.1)$) CKM phase if $\langle T \rangle$
is the only source of CP violation.
We note that the dilaton also has a complex VEV in order to produce a
relative sign between  the condensates and  Im$S$ is fixed up to a discrete shift
\cite{deCarlos:1993da}.

In all interesting cases, $F_T$ receives a complex phase of
order one. This occurs due to a rapid variation of $G_T$
(Eq.\ref{G}) around the fixed points. However, $F_T$ is not
modular invariant by itself so this does not necessarily mean that
the physical SUSY CP phases are also ${\cal O}(1)$. We will
discuss this subject separately in one of the subsequent sections.

Introduction of the Green-Schwarz term does not significantly change
the situation, as should probably be expected. 
The only relevant change is that now $F_S$ differs from zero due to
the dilaton-modulus mixing (Table \ref{table2}).
In all cases $F_S $ is of order $F_T$ and also has an order one complex phase.

\subsection{Models With Non-Perturbative Corrections to the
K\"{a}hler Potential}

One generally expects the  K\"{a}hler potential to receive corrections from
non-perturbative effects. Such effects may be responsible for dilaton stabilization.
For instance,
in Ref.\cite{Casas:1996zi} the following  K\"{a}hler potential was suggested:
\begin{equation}\label{nonperk}
  K_S=\ln \left(\frac{1}{2 \mathrm{Re}
  S}+d(\mathrm{Re}S)^{\frac{p}{2}}e^{-b \sqrt{\mathrm{Re}S}}\right),
\end{equation}
where $d,p,b$ are certain constants ($p,b >0$).

\begin{figure}[ht]
\begin{center}
\begin{tabular}{c}

\epsfig{file=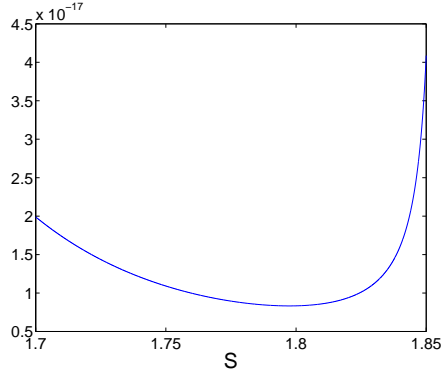, height=5cm}

\end{tabular}
\end{center}
\caption{  Scalar potential with non-perturbative K\"{a}hler
potential, $m=n=0$. $T$ is set to its minimum value, $e^{\pm
i\pi/6}$. The minimum in $S$ is at $1.8$. } \label{CPfig5}
\end{figure}

\begin{figure}[ht]
\begin{center}
\begin{tabular}{c}

\epsfig{file=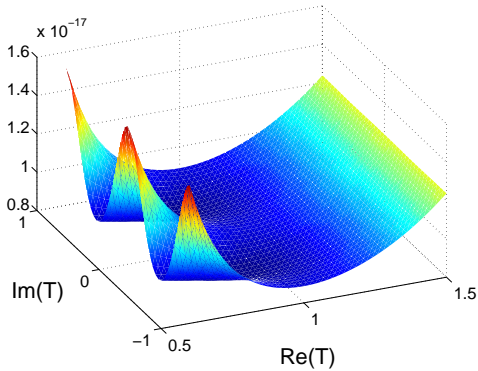, height=5cm}

\end{tabular}
\end{center}
\caption{ Scalar potential with non-perturbative K\"{a}hler
potential, $m=n=0$. S is set to its minimum value, 1.8. The minima
in $T$ are at $e^{\pm i\pi/6}$. Note the invariance of the
potential under the axionic shift $T\rightarrow T+i$.}
\label{CPfig6}
\end{figure}
The superpotential remains given by Eq.\ref{trunW}.
The second term under the log in Eq.\ref{nonperk} represents non-perturbative
corrections. Its form based on the natural assumptions that non-perturbative
effects vanish in the limit of vanishing coupling constant (${\rm Re}S\rightarrow \infty
$) and are zero at all orders of perturbation theory.
Note that the scalar potential is a function of
Re$S$ only, so  Im$S$ remains undetermined.

In Table \ref{table3} we present the results of the potential minimization
for $m$ and $n$ between 0 and 5 with $d=7.8$ and $b=p=1$.
The presence of $H(T)$ does not visibly affect the value of $S$ at the minimum, but
typically forces $T_{\rm min}$  into the fixed points.
At these points no CP violation is produced and
SUSY remains unbroken unless $m=n=0$ (recall that
$H(e^{\pm i \pi/6})\bigl\vert_{n>0}= H(1)\bigl\vert_{m>0}= 0$).
Clearly, these minima are phenomenologically unacceptable.
Having varied $p,d,b$, we were unable to find viable CP-violating minima
with a reasonable SUSY breaking scale. The same remains true for a nonzero
$\delta_{GS}$ (Table \ref{table4}).

This however does not mean that this stabilization mechanism is altogether
unattractive. It has a nice feature that  dilaton-dominated SUSY
breaking can be obtained and most of the soft CP phases can be eliminated. 
CP violation in this case may originate
from fields other than the dilaton and moduli, for instance,  
Froggatt-Nielsen type fields. To avoid the SUSY CP problem, one needs to ensure that such fields do not break supersymmetry. 

\subsection{Models With S-Dual Potentials}

\begin{figure}[ht]
\begin{center}
\begin{tabular}{c}
\epsfig{file=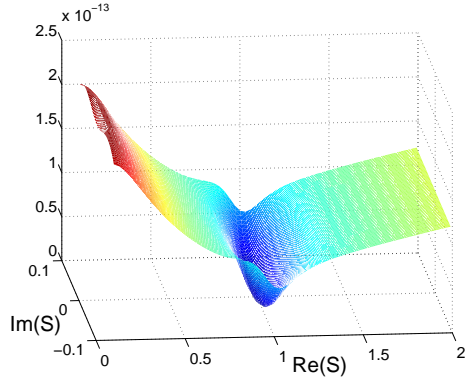, height=5cm}
\end{tabular}
\end{center}

\caption{S-dual scalar potential with $H$ and m=1, n=0. $T$ is set
to its minimum value, $0.9850e^{0.5471i}$. The minimum in $S$ is
at $S_{min}=1$. }

\label{CPfig3}
\end{figure}


\begin{figure}[ht]
\begin{center}
\begin{tabular}{c}

\epsfig{file=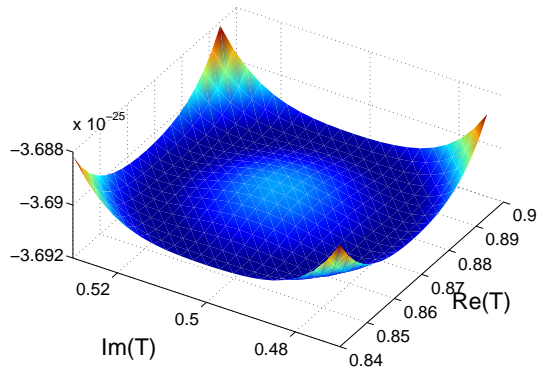, height=5cm}

\end{tabular}
\end{center}
\caption{ S-dual scalar potential with $H$ and m=1, n=0. $S$ is
set to its minimum value, 1. The minimum in $T$ is at
$T_{min}=0.9850e^{0.5471i}$. } \label{CPfig4}
\end{figure}
It is feasible that the underlying theory can possess $SL(2,{\bf {\rm Z}})$
S-invariance in addition to the well known T -- modular invariance.
S-self-dual models naturally exhibit dilaton stabilization as large and small
values of S are related by S-duality \cite{Lalak:1995hn}.
In general, the gauge coupling is given
by the gauge kinetic function $f$:
\begin{equation}
{\rm Re }f=1/g^2\;,
\end{equation}
so one can have $g \rightarrow g$ under $S\rightarrow 1/S$ if $f \rightarrow f$,
as opposed to the standard strong--weak duality. Since in the weak coupling
limit (${\rm Re}S \rightarrow \infty$) $f(S) \rightarrow S$, the simplest
S-invariant kinetic function is \cite{Lalak:1995hn}
\begin{equation}\label{Sf}
 f_s=\frac{1}{2 \pi} \ln \left( j(S)-744 \right)\;.
\end{equation}
Note that this duality relates theories with the same
coupling, but differing values of $S$. The standard K\"{a}hler potential
$ K=-\ln (S+ \bar S)-3\ln(T+\bar {T} -X \bar X) $, where $X^3=W^\alpha W_\alpha$, then
implies that the superpotential must have weight -1 with respect to the S-duality
transformation:
\begin{equation}\label{sdualw}
 W=\frac{X^3}{ \eta^2 (S)}\left[ \frac{1}{2 \pi} \ln (j(S)-744)+ 3 b
 \ln (X \eta^2(T)/\mu) +c \right]\;,
\end{equation}\
where $b=2\beta/3$ and  $\mu , c$ are constants. Integrating out
the condensate using the truncated approximation, we get, after absorbing the
constant $c$ into $\mu$,
\begin{equation}\label{sdutrw}
 W=-\frac{2 \beta \mu^3}{3e \eta(S)^2 \eta(T)^6 (j(S)-744)^{\frac{3}{4 \pi
 \beta}}}\;.
\end{equation}

 In our numerical analysis, we consider a model with a $SU(6)$
gauge group. The corresponding minima for the case without matter
are  shown in Table \ref{table5}. The situation is very similar to
the case of the racetrack models apart from the value of $S_{\rm
min}$. $F_S$ always vanishes since  $S$ is stabilized at the fixed
point. 
For  $m\geq 1$, $n>0$, the minima in $T$ are located on the unit circle
where $F_T$ vanishes.   
The only reasonable minimum appears  for
$m=1,n=0$, but again it is close to the fixed point where the CKM
phase vanishes. Figs.\ref{CPfig3} and \ref{CPfig4} illustrate the
behaviour of the potential, which is very similar to what we have
seen in the racetrack models.

We do not consider $\delta_{GS}\not= 0 $ case since it is not clear
whether one can maintain both T- and S- modular invariance at the
one loop level.

\section{The CKM Phase }

In this section we will briefly discuss how the CKM phase can be produced
in heterotic orbifold models. One of the crucial requirements
any model should satisfy is that it reproduces the standard CKM
picture of CP violation and the consequent CKM phase is of order one.   

Complex Yukawa couplings in heterotic string models can be generated
if the matter multiplets belong to the twisted sectors and the moduli
fields attain complex VEVs.
In general, this does not necessarily mean that a nonzero CKM
 phase is produced. The complex phases in the
Yukawa matrices can often be removed by a basis transformation
consistent with the $SU(2) \times U(1)$ symmetry. The proper measure
of CP violation in the Standard Model is given by the Jarlskog invariant
\cite{Jarlskog:1985ht}:
\begin{eqnarray}
J&=&{\rm Im~\biggl(~ det} \left[ Y^u Y^{u\dagger}, Y^d Y^{d\dagger}  \right]~
\biggr)\;,
\end{eqnarray}
where $Y^{u,d}$ are the Yukawa matrices. A non-zero $J$ indicates the
presence of the CKM phase.

The (renormalizable) Yukawa couplings can be calculated exactly in a
given heterotic orbifold model. Often the Yukawa couplings have a very
restricted flavour structure such that the complex phases are spurious
and the Jarlskog invariant vanishes. That is the case for the prime order
orbifolds, whereas for the non-prime orbifolds the CKM phase can be
non-trivial \cite{Lebedev:2001qg}. Here we will give an example of
the $Z_6$-I orbifold
assuming that we  have the freedom to assign a field to a fixed point
of our choice \cite{Casas:1990qx,Casas:1993ac}.
Note however that we do not attempt to reproduce the observed fermion
masses and mixings, so this picture is not fully realistic. Nevertheless,
it gives a fair idea of CP violation in the system.

Due to string selection rules, only fields belonging to particular fixed
points can couple via the Yukawa interaction. This restricts the flavour
structure of the Yukawa matrices.  In the $Z_6$-I case, one of the allowed
couplings is $\theta \theta^2 \theta^3$, where $\theta^i$ denote the twisted
sectors.
The corresponding $f_1 f_2 f_3$ Yukawa couplings are expressed as
\cite{Casas:1993ac}
\begin{eqnarray}
&& Y_{\theta \theta^2 \theta^3}=
N \sqrt{l_2 l_3} \sum_{\stackrel{\rightarrow}{u} \in Z^4}
{\rm exp} \biggl[ -4\pi T \left(
\stackrel{\rightarrow}{f_{23}} + \stackrel{\rightarrow}{u} \right)^T
M \left(
\stackrel{\rightarrow}{f_{23}} + \stackrel{\rightarrow}{u} \right)
\biggr]\;,
\label{yukawa}
\end{eqnarray}
where $f_{1,2,3}$ are the fixed points,
$\stackrel{\rightarrow}{f_{23}}$ represents a
projection of $f_2-f_3$ onto the first two complex planes (corresponding
to $T_1$ and $T_2$),
$\stackrel{\rightarrow}{u}$ is a four-dimensional vector with integer
components, $N$ is a normalization factor,
$l_i$ are the ``multiplicity'' constants associated with the fixed points,
 and the matrix $M$ is
given by
\begin{eqnarray}
&& M=\left( \matrix{1 & -{3\over2} & 0 & 0 \cr
                    -{3\over2} & 3 & 0 & 0 \cr
                    0 & 0 & 1 & -{3\over2} \cr
                    0 & 0 & -{3\over2} & 3} \right)\;.
\end{eqnarray}
Clearly, this Yukawa coupling is complex for complex $T$.
Next, we need to assign the observable fields to the fixed points.
One possible assignment producing a nonzero Jarlskog invariant is
given in Table \ref{z6}.
\begin{table}
\begin{center}
\begin{tabular}{|c||c|c|}
\hline
  {\rm field}   &   {\rm fixed point} & $l$  \\
\hline
\hline
$H_{1,2}$ & $(0,0)\otimes (0,0) $ & 1 \\
$Q_1$ & $(0,0)\otimes (0,0) $ & 1 \\
$Q_2$ & $\left( 0,{1\over3} \right) \otimes \left( 0,{1\over3} \right) $ & 2 \\
$Q_3$ & $\left( 0,{1\over3} \right) \otimes \left( 0,0 \right) $ & 2 \\
$U_1$ & $(0,0)\otimes (0,0) $ & 1 \\
$U_2$ & $\left( 0,{1\over2} \right) \otimes \left( 0,{1\over2} \right) $ & 3 \\
$U_3$ & $\left( {1\over2},{1\over2} \right) \otimes \left( 0,{1\over2} \right) $ & 3 \\
$D_1$ & $\left( 0,{1\over2} \right) \otimes \left( 0, 0 \right) $ & 3 \\
$D_2$ & $\left( 0,0 \right) \otimes \left( 0,{1\over2} \right) $ & 3 \\
$D_3$ & $(0,0)\otimes (0,0) $ & 1 \\
\hline
\end{tabular}
\end{center}
\caption{$Z_6$-I fixed point assignment for the observable fields. }
\label{z6}
\end{table}
Here the Higgs fields are assumed to belong to the $\theta$ sector,
quark doublets -- to the $\theta^2$ sector, and the quark singlets
-- to the $\theta^3$ sector.
We note that above the electroweak symmetry breaking scale the quark
fields may also appear as linear combinations of the ones in Table \ref{z6},
this however does not affect the Jarlskog invariant.
Finally, the Yukawa couplings
must be rescaled $ Y_{abc}\rightarrow Y_{abc} {\hat W^*}/\vert {\hat W} \vert e^{\hat K/2}
(K_a K_b K_c)^{-1/2}$ \cite{Brignole:1997dp} in order to have the canonical
normalization  and to be  weight zero
quantities under the modular transformation.

\begin{figure}[ht]
\begin{center}
\begin{tabular}{c}
\epsfig{file=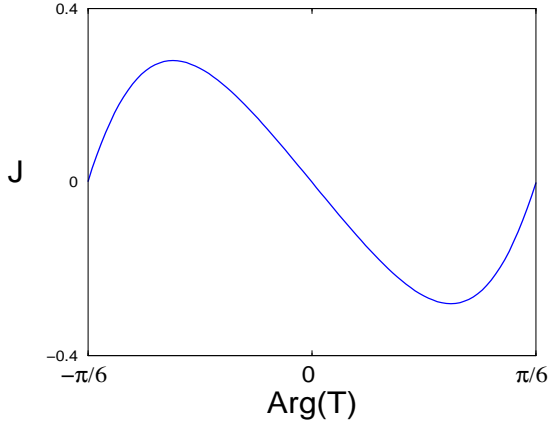, width=7.4cm, height=5.8cm}\\
\end{tabular}
\end{center}
\caption{Jarlskog invariant for the $Z_6$-I orbifold as a function of  ${\rm Arg}( T )$ 
for $T$ on the unit circle.
 }
\label{jar}
\end{figure}

The corresponding Jarlskog invariant as a function of the modulus field $T$
is presented in Fig.\ref{jar}. We restrict $T$ to be on the unit circle which is
often the case of interest. The overall normalization of $J$ is irrelevant for our
purposes since we are not producing the observed quark mass hierarchy
and mixings, however the figure provides important qualitative features.
In particular, $J$ vanishes at the fixed points of the modular group
$1$, $\exp (\pm i \pi/6)$ due to the axionic shift invariance
\cite{Lebedev:2001qg}.
Away from the fixed points it is non-zero. If the Standard Model sector
exhibited the T-duality invariance, the CKM phase would have to vanish
everywhere on the unit circle \cite{Dent:2001cc}. However, typically the SM sector
interactions are not modular invariant\footnote{The duality symmetry is broken
spontaneously at high energies and can only be non-linearly realized at
the electroweak scale.}.
This can be seen directly from the action
of the duality transform on the fields at the fixed points
\cite{Lebedev:2001qg}.
The fields necessary to restore full modular
invariance  are associated with heavy matter fields and decouple at low
energies. This situation is  analogous to what we encounter
in GUT models, say  $E_6$. The low energy spectrum does not
form a representation of $E_6$
and to restore the symmetry one has to add extra heavy fields.

To conclude this section, we have argued that generally it is possible
to generate a CKM phase
at the renormalizable level
through a complex VEV of the modulus field (away from the fixed points).
For instance, 
order one CKM phase can be produced with Arg$T\sim {\cal O}(\pi/12)$
for $T$ on the unit circle.
Naturally, we expect that the   CKM phase can be induced  in a larger
class of models if nonrenormalizable operators are taken into account
\footnote{See, for instance, \cite{Faraggi:1994su} for a related discussion. }.

\section{Soft SUSY Breaking Terms}

In this section we will consider soft SUSY breaking terms for the minima
obtained in the previous sections. The purpose of our analysis is to establish
how much CP violation in the soft terms should generally  be expected
if both dilaton and moduli are  stabilized by the underlying dynamics.

Before we list the  formulae for the soft breaking terms, let us make explicit our
notation. The soft SUSY breaking Lagrangian is given by
\begin{eqnarray}
{\cal L}_{\rm soft}={1\over 2}\left( M_a \lambda^a \lambda^a +{\rm h.c.} \right)
- m_\alpha^2 \hat\phi^{*\alpha}  \hat\phi^{\alpha}-\biggl(  {1\over 6}
 A_{\alpha \beta \gamma}  \hat Y_{\alpha \beta \gamma} \hat\phi^\alpha  \hat\phi^\beta
\hat\phi^\gamma + B \hat \mu \hat H_1 \hat H_2 +{\rm h.c.} \biggr)\;,
\end{eqnarray}
where $\hat Y_{\alpha \beta \gamma}$ and $\hat \mu$ are  the Yukawa couplings and the
$\mu$-term
for the  canonically normalized fields $\hat\phi$. 
With the K\"{a}hler potential and the superpotential
of the form
\begin{eqnarray}
&& K= \hat K +\tilde K_\alpha \phi^{*\alpha}  \phi^{\alpha}
+\left(  Z H_1 H_2 + {\rm h.c.} \right)\;, \nonumber\\
&&W=\hat W +  {1\over 6} Y_{\alpha \beta \gamma} \phi^\alpha  \phi^\beta
\phi^\gamma  \;,
\end{eqnarray}
$\hat Y_{\alpha \beta \gamma}$ and $\hat \mu$ are given by \cite{Brignole:1997dp}
\begin{eqnarray}
&&\hat Y_{\alpha \beta \gamma}=Y_{\alpha \beta \gamma} {\hat W^* \over \vert \hat W \vert }
e^{\hat K/2} \left( \tilde K_\alpha  \tilde K_\beta  \tilde K_\gamma \right)^{-1/2} \;,
\nonumber\\
&&\hat \mu = \left( m_{3/2} Z- \bar F^{\bar m} \partial_{\bar m} Z\right)
\left( \tilde K_{H_1} \tilde K_{H_2}\right)^{-1/2}\;.
\end{eqnarray}
Here $m=(S,T)$ and for definiteness we have assumed the Giudice-Masiero mechanism for
generating the $\mu$-term
\cite{Giudice:1988yz}.
The canonically normalized fields are obtained by the rescaling 
$\hat \phi_\alpha = \tilde K_\alpha^{1/2} \phi_\alpha$.
The gaugino masses, scalar masses, A-terms, and the B-term are expressed,
respectively,
as \cite{Brignole:1997dp}:
\begin{eqnarray}
\label{softterms}
 M_a &=& {1\over 2}({\rm Re}f_a)^{-1} F^m \partial_m f_a\;,  \\
 m_\alpha^2 &=& m^2_{3/2}+V_0 - \bar F^{\bar m} F^n \partial_{\bar m} \partial_n
\ln \tilde K_\alpha\;,\nonumber\\
 A_{\alpha \beta \gamma} &=& F^m \left[ \hat K_m +\partial_m \ln Y_{\alpha \beta \gamma}
-\partial_m \ln (\tilde K_\alpha \tilde K_\beta \tilde K_\gamma)
\right]\;,\nonumber\\
 B &=& \hat \mu^{-1} \left(\tilde K_{H_1} \tilde K_{H_2}\right)^{-1/2} \biggl[
(2 m^2_{3/2}+V_0) Z - m_{3/2} \bar F^{\bar m} \partial_{\bar m} Z
\nonumber\\
&+& m_{3/2} F^m \left( \partial_m Z-Z ~\partial_m \ln(\tilde K_{H_1} \tilde K_{H_2})  \right)
-\bar F^{\bar m} F^n \biggl( \partial_{\bar m} \partial_n Z
- \partial_{\bar m} Z~\partial_n \ln(\tilde K_{H_1} \tilde K_{H_2})  \biggr)
\biggr]\;.\nonumber
\end{eqnarray}

Note that the gaugino masses computed with the kinetic
function (\ref{ff}) do not appear to be modular invariant. An additional contribution from
the massless fields of the theory is necessary to rectify this problem. Effectively
this amounts to an addition of the non-holomorphic term
$2 \beta \ln(T+\bar T)$ to the kinetic function \cite{Ibanez:1992hc}.

Before we proceed let us clarify our framework. In what follows, we will not
restrict ourselves to a particular orbifold model. Instead, we will try to present
some general features  of models possessing modular invariance.
At the same time, we will clarify some of our statements with explicit examples.

The K\"{a}hler function $\tilde K_\alpha$ is expressed as
\begin{equation}
\tilde K_\alpha=(T+\bar T)^{n_\alpha} \;,
\end{equation}
where $n_\alpha$ is a modular weight.
Here we have assumed a diagonal K\"{a}hler metric which is almost always the case in
phenomenologically acceptable models.
The reason is that the space group quantum numbers typically prohibit
an off-diagonal metric  \cite{Brignole:1997fb}.
Moreover, our main results are unaffected even if we allow for a mixing among the fields
belonging to the same twisted sector.

For non-oscillator states, we have
\cite{Ibanez:1992hc}
\begin{eqnarray}
&& n_\alpha^{\rm untw.}=-1\;, \nonumber\\
&& n_\alpha^{\rm tw.}= -2  \;\;({\rm three\; planes\; rotated})\;,\nonumber\\
&& n_\alpha^{\rm tw.}= -1  \;\;({\rm two\; planes\; rotated})\;.
\end{eqnarray}
where we distinguished between the possibilities with the twists
in all planes being nonzero and the twists in two planes
being nonzero. The oscillator states usually appear as singlets
and are not associated with the MSSM fields \cite{Brignole:1997fb},
so we will restrict our discussion to the non-oscillator states only.

It is interesting to note that in order to get nontrivial Yukawa textures and
the CKM phase, the  MSSM fields  associated with different generations should
 belong to the same twisted sector. For instance, in our $Z_6$-I example all
quark doublets belong to the $\theta^2$ sector, whereas all quark singlets
are in the $\theta^3$ sector. If, say $Q_1$, belonged to the $\theta$ or
$\theta^3$ sector, its coupling with $H_2 U_i$ or $H_1 D_i$ would be prohibited
since only the coupling of the form $\theta \theta^2 \theta^3$  is allowed
(if the Higgses are fixed to be in the $\theta$ sector). This would result
in the Yukawa textures containing many zeros and the Jarlskog invariant
would be likely to vanish.

This observation has implications for the Flavour Changing Neutral Currents (FCNC)
since it implies that the modular weights and
thus $m_\alpha^2$ are $generation-independent$. Excessive FCNC at low energies result
from non-degeneracy of the squark masses and generally pose a problem for supersymmetric
model building \cite{Gabbiani:1996hi}.
Here it is naturally avoided if we are to produce the CKM phase\footnote{A flavour dependence in
$A_{\alpha \beta \gamma}$ can also contribute to FCNC, however it generates left-right
squark mass insertions proportional to the quark masses
which are only loosely  constrained \cite{Gabbiani:1996hi}.}.

Similarly, in the A-terms   the only generation-dependent piece  comes from the Yukawas
($\tilde K_\alpha \tilde K_\beta \tilde K_\gamma$ is fixed). Nevertheless, this dependence
can be strong because  the  term  $\partial_m \ln Y_{\alpha \beta \gamma} $
can be significant and even dominant. In particular, it is easy to see from Eq.\ref{yukawa}
that for ${\rm Re} T \sim 1$, the Yukawa coupling is dominated by one term.
Consequently,
\begin{equation}
\partial_T \ln Y_{\alpha \beta \gamma} \simeq -4\pi
\stackrel{\rightarrow}{h_{23}}^{ T} M \stackrel{\rightarrow}{h_{23} }
\label{lnY}
\end{equation}
 for some (typically fractional) $\stackrel{\rightarrow}{h_{23}}$
depending  on  $\alpha,\beta,\gamma$.
Independently of T, $\partial_T \ln Y_{\alpha \beta \gamma} $
is an almost real number typically
between -1 and -10 (for $\stackrel{\rightarrow}{h_{23}}=0$ it is zero).
This creates a significant flavour dependence in the A-terms and thus a flavour universality
of the minimal SUGRA model cannot be achieved.
Note that if  the correct fermion mass hierarchy is reproduced, this effect may become even
stronger.  These conclusions equally apply to other orbifolds.

If the MSSM fields are in the untwisted sector, the soft terms (apart from the gaugino masses)
 are universal. This possibility however is unattractive since the Yukawa couplings are either
zero or one. In this case there is no fermion mass hierarchy and the CKM phase has to vanish.
 Even if non-renormalizable operators are taken
into account this option is hardly  phenomenologically viable
\cite{Casas:1990qx}.

The CP-violating phases appearing in the $B$- and $\mu$- terms critically depend on
the $\mu$-term generation mechanism. The ``bare'' $\mu$ parameter appearing
in the superpotential would have to be of order Planck scale which is phenomenologically
unacceptable. Thus a different mechanism is required. One of the attractive  ways
to produce the $\mu$-term of order $m_{3/2}$ is the Giudice-Masiero mechanism
\cite{Giudice:1988yz}.
This mechanism employs the K\"{a}hler symmetry of the theory so that a  K\"{a}hler
transformation induces an effective $\mu$-term in the Lagrangian even though
it was not present initially. This requires the presence of the $ZH_1H_2$ term
in the   K\"{a}hler potential, 
which can be implemented in string models \cite{Antoniadis:1994hg}.
Such a term arises in even order orbifolds possessing  at least one complex structure
modulus $U$.  Specifically, in these models $Z$ has the form
\begin{equation}
Z={1\over (T_3+T_3^*)(U_3+U_3^*)}\;,
\end{equation}
where $T_3$ is associated with the $Z_2$ plane of the orbifold
and the Higgses are assumed to be untwisted. One can check that the
K\"{a}hler potential has proper transformation properties
up to ${\cal O}\Bigl((H_1H_2)^2\Bigr)$ under T-duality
if the U-modulus transforms as $U\rightarrow U -H_1 H_2 ic/(icT+d)$
\cite{Antoniadis:1994hg}.

\subsection{SUSY CP Phases and Modular Invariance}

In this subsection we will address the question of modular invariance of the physical
CP phases. These phases are invariant under the $U(1)_R$ and $U(1)_{PQ}$ and are given by
\cite{Abel:2001vy}:
\begin{equation}
{\rm Arg}\biggl( (B\hat\mu)^*\hat\mu M \biggr)\;,\;{\rm Arg}\left( A^*M \right)\;,
\label{phases}
\end{equation}
where the A-terms and the gaugino masses are assumed to have universal phases.
For clarity of our presentation we will assume $\delta_{GS}=0$.
Also, as we have seen above, in all relevant cases $F_S=0$, so henceforth
we will set $F_S$ to zero.
In what follows, we will consider in detail only the duality
transformation $T\rightarrow 1/T$ since the discussion of the axionic shift invariance
is quite trivial.

Let us first define $strictly$ weight 2 modular functions 
(related to the Eisenstein function)
\begin{eqnarray}
&& G_2(T)=  {1\over T+\bar T}+ 2{\eta'(T) \over \eta(T)}\;,\nonumber\\
&&\tilde G_2(T)=  {1\over T+\bar T}+ 2{\eta'(T) \over \eta(T)}-{1\over 3}
{H'(T) \over H(T)} \;.
\end{eqnarray}
They indeed transform with a modular weight +2  since they are given by a logarithmic
derivative of $(T+\bar T)\eta^2(T)$ and  $(T+\bar T)\eta^2(T)H(T)^{-1/3}$.
The auxiliary field $F^T$ is then given by
\begin{equation}
F^T=(T+\bar T)^2 \tilde G_2(T)^* \times {\rm modular\; invariant\; piece}\;.
\end{equation}
Since under duality $T+\bar T \rightarrow (T+\bar T)/T\bar T$ and $\tilde G_2
\rightarrow -T^2 \tilde G_2~$, $F^T$ transforms as
\begin{equation}
F^T \rightarrow -{1\over T^2}~F^T\;.
\label{f}
\end{equation}

Let us now consider how $\hat\mu$ and $B\hat\mu$ transform under duality. For $U+\bar U=1$ and
$F^U=0$, we have
\begin{eqnarray}
&&\hat\mu=m_{3/2}+ {{\bar F^T} \over T+\bar T}\;, \nonumber\\
&&B\hat\mu=2 m_{3/2}^2+V_0+m_{3/2} {F^T+{\bar F^T} \over T+\bar T}\;.
\end{eqnarray}
These expressions are apparently modular non-invariant. However, one must keep in mind that
the U-modulus is not inert under duality. It provides the necessary terms to restore
modular invariance. Indeed, the relevant terms in the K\"{a}hler potential are
\begin{equation}
\Delta K=-\ln(U+\bar U) +{H_1 H_2 +{\rm h.c.}\over (T+ \bar T)(U+ \bar U)}\;.
\end{equation}
Under duality $U \rightarrow U - H_1 H_2/T$
\cite{Antoniadis:1994hg} and $\Delta K $ remains invariant up to
${\cal O}(H_1^2H_2^2)$ terms  which are consistently neglected in
supergravity. In terms of the function $Z(T,U)$, this translates
into the following ``anomalous'' transformation property
\begin{equation}
Z\rightarrow ZT \bar T-T\;,
\end{equation}
where we have set $U+ \bar U=1$.
Using this fact and recalling that the Higgses are untwisted, one can show that
under duality
\begin{eqnarray}
&&\hat \mu \rightarrow -{T\over \bar T} \hat \mu\;, \nonumber \\
&&B\hat \mu \rightarrow -{T\over \bar T} B\hat \mu\;.
\end{eqnarray}
One may be wondering whether such transformations with a T-dependent phase are consistent with
modular invariance of the theory. To answer this question one should recall that
the canonically normalized untwisted fields transform under duality as
\begin{equation}
\hat \phi \equiv \phi (T+\bar T)^{-1/2}
\rightarrow  -i \left({\bar T\over  T}\right)^{1/2}  \hat \phi\;.
\end{equation}
As a result, the interaction terms  $\hat \mu \hat H_1 \hat H_2$ and
$B\hat \mu \hat H_1 \hat H_2$ are $exactly$ modular invariant. The same result can be
established for the $\mu$-term generated non-perturbatively.
We therefore see that Arg$((B\hat \mu)^* \hat\mu)$ 
is modular invariant.

The gaugino masses calculated with the kinetic function augmented by  $2 \beta \ln(T+\bar T)$
\cite{Ibanez:1992hc} are modular invariant. Indeed,
\begin{equation}
M_a \propto F^T G_2(T) \rightarrow M_a
\end{equation}
since  $F^T$ transforms with modular weight -2 (Eq.\ref{f}). Thus the phase of $M_a$ is
modular invariant. Note that the gaugino masses have a universal phase due to $F^S=0$,
whereas their magnitudes are proportional to the beta functions and thus are different.

The discussion of the A-terms is more involved. The complication comes from the term
$\partial_T \ln Y_{\alpha\beta\gamma}$; as we know the Yukawas couplings have  highly
non-trivial transformation properties under duality. In fact, if we associate the MSSM
fields with a subset of the fixed points of the orbifold, generally this subset does not
transform into itself under duality \cite{Lebedev:2001qg}. Consequently, the Standard Model
interactions are generally non-invariant under duality, although the full set
associated with all of the fixed points is invariant.
So leaving aside these flavour issues, the best we can do is to study the overall phase
of the A-terms. As a matter of fact, this is a very good approximation
for Re$T\simeq 1$.
The reason is that the Yukawa couplings are dominated
by one term, so   $\partial_T \ln Y_{\alpha\beta\gamma}$ is real to a very
good degree (Eq.\ref{lnY}). The other relevant terms
$\hat K_m$  and $\partial_m \ln (\tilde K_\alpha \tilde K_\beta \tilde K_\gamma)$
are also real, so $A_{\alpha\beta\gamma}$ has a universal phase up to small corrections,
although its  magnitude can be highly non-universal.
This property remains valid under a duality transformation since again the sum is dominated
by one term (if Re$T\simeq 1$); this can also be seen from the expressions for the
duality-transformed Yukawas \cite{Lebedev:2001qg}.
We have checked numerically that the deviations from universality are within a few percent
(apart from the suppressed elements $A_{\alpha\beta\gamma}\simeq0$)
for the $Z_6$-I model with $T$ close to the unit circle.
So for practical purposes we
can treat ${\rm Arg}(A)$ as a universal phase.

Denoting by $n_{\alpha}$ a modular weight of the relevant field, the A-terms can be cast in
the following form
\begin{eqnarray}
&& A_{\alpha\beta\gamma}=F^T \partial_T  \ln \biggl[
Y_{\alpha\beta\gamma} (T+\bar T)^{-3-n_\alpha-n_\beta-n_\gamma}
\biggr]\;.
\end{eqnarray}
The modular weight of the Yukawa coupling is fixed by requiring
the superpotential to transform with modular weight -3. So apart
from the unitary transformation mixing the fields in each twisted
sector, we have
\begin{equation}
Y_{\alpha\beta\gamma} \rightarrow (iT)^{-3-n_\alpha-n_\beta-n_\gamma} Y_{\alpha\beta\gamma}\;.
\end{equation}
It is easy to see that the A-terms stay invariant under duality,
\begin{equation}
A_{\alpha\beta\gamma} \rightarrow A_{\alpha\beta\gamma}\;.
\end{equation}
The corresponding interaction term $A_{\alpha\beta\gamma} \hat Y_{\alpha\beta\gamma}~
\hat \phi_\alpha \hat \phi_\beta \hat \phi_\gamma $ also stays invariant if we recall
\begin{eqnarray}
 \hat \phi_\alpha &\rightarrow& i^{n_\alpha} \left( {T\over \bar T}\right)^{n_\alpha/2}
\hat\phi_\alpha\;,
\nonumber\\
 (\tilde K_\alpha \tilde K_\beta \tilde K_\gamma )^{-1/2} &\rightarrow&
(T \bar T)^{(n_\alpha+n_\beta+n_\gamma)/2} ~
(\tilde K_\alpha \tilde K_\beta \tilde K_\gamma )^{-1/2}\;,\;\nonumber\\
 e^{\hat K/2} &\rightarrow& (T \bar T)^{3/2} e^{\hat K/2} \;, \nonumber\\
 {\hat W^* \over \vert \hat W \vert } &\rightarrow& (-i)^{-3}
\left( {T \over \bar T}\right)^{3/2} {\hat W^* \over \vert \hat W \vert }\;.
\end{eqnarray}
Here we have used the fact that $\hat W$ transforms with weight
-3. In practice this is true only up to a T-independent phase
which can always be absorbed into redefinition of the fields. One
should keep in mind that here we have ignored the unitary
transformation in the twisted sector which accompanies the duality
transformation, but this is justified as long as we are concerned
with the overall phase of the A-terms.

The discussion simplifies if matter is untwisted. In this case the A-terms vanish
and the requirement of modular invariance is trivially satisfied.

To summarize the results of this section,
we find that even though the individual CP phases may not be modular invariant,
the physical CP phases of Eq.\ref{phases} are (at least under our assumptions).

Before we proceed to the numerical analysis, let us make an important comment.
We would like to stress that  there can be no CP violation induced by a VEV
of the modulus field if it is stabilized at the fixed point,
at least for $\delta_{GS}=0$.
The Jarlskog invariant vanishes at the fixed points regardless of the
presence of the Green-Schwarz term, so
there is no CKM phase. For  $\delta_{GS}=0$, $F_T$ vanishes at the fixed points,
so no soft phases are induced.
Further, flavour-dependent complex phases in
$A_{\alpha \beta \gamma}  \hat Y_{\alpha \beta \gamma}$ arising from the phases
in the Yukawa matrix can be removed by a phase redefinition of
the quark superfields\footnote{This can be seen explicitly from the phase factorization
properties of the Yukawas (under $T\rightarrow T+i/2$) of  Ref.\cite{Lebedev:2001qg}.}.
In principle, CP violation can  be induced by complex $S$ and $F_S$ but the dilaton
does not distinguish flavours, so even in this case the CKM phase is zero.
Thus, the conclusion is that no realistic CP violation can be produced for
$T$ at the fixed points.
This of course is also true if $T$ is sufficiently close to the fixed points.
In our case $T=0.985~ e^{0.5417 i}$ which is close to $e^{i\pi/6}$. The resulting
$non-removable$ phases in the Yukawa matrix are of order $10^{-1}$, so the CKM
phase is suppressed.

\subsection{Numerical Results }

In all interesting cases the modulus field is stabilized close to the fixed points.
As we know, supersymmetry is unbroken at the fixed points, so $F_T$ takes on
a rather small value compared to $m_{3/2}$ for $T \simeq e^{i\pi /6}$.
This leads to the problem of tachyons (see, e.g. \cite{Bailin:1999nk}).
Indeed, since
\begin{equation}
V_0 \sim  -3m_{3/2}^2\;,
\end{equation}
the soft sfermion masses in Eq.\ref{softterms} are dominated by the $V_0$ term and
\begin{equation}
m_{\alpha}^2 \sim -2 m_{3/2}^2\;.
\end{equation}
This is of course a problem. One may impose the condition of the vanishing
cosmological constant to begin with, but it would be extremely difficult
to obtain dilaton stabilization, CP violation, and correct SUSY breaking
at the same time. For the lack of a better solution, we may simply assume
the there is an additional contribution to the K\"{a}hler potential,
$K(X,\bar X)$, which allows us to set $V_0$ to zero \cite{Bailin:1999nk}.
This contribution will have an effect on all of the soft terms other than the gaugino
masses and will help avoid tachyons. Another possibility is to include 
the effect of quantum corrections \cite{Choi:1994xg}.

Another problem arises from the gaugino masses. In the absence of the dilaton
SUSY breaking, the gaugino masses are suppressed by a loop factor $\beta$.
In addition to that, they are proportional to $G_2(T)$ which is suppressed
close to a fixed point. This results in a suppression factor of about
$10^3$. The problem is ameliorated in the presence of the Green-Schwarz term
(for multiple gaugino condensates) which creates a non-zero $F_S$
\cite{deCarlos:1992pd}.

Concerning the magnitudes of the soft terms, for 
a representative point  
$T=0.985~ e^{0.5417 i}$ (racetrack  $\delta_{GS}=0$ and S-dual models,
$m=1, n=0$),
  we obtain
\begin{eqnarray}
 M_a &\sim& 10^{-1}-1\;GeV \;,\nonumber\\
 m_\alpha &\sim& 10^4\;GeV \;\;({\rm tachyonic})\;,\nonumber\\
 A_{\alpha\beta\gamma} &\sim& 10^3\;GeV \;,\nonumber\\
 \hat \mu &\sim& 10^4\; GeV \;,\nonumber\\
\sqrt{ B \hat \mu} &\sim& 10^4 \; GeV\;.
\end{eqnarray}
This SUSY spectrum as it stands is of course phenomenologically unacceptable.
Significant modifications of the model are necessary.  One possibility
would be a mechanism producing a substantial dilaton SUSY breaking component,
$F_S \not =0$. This would certainly rectify the problem of light gauginos
and help avoid tachyons. 
The presence of the Green-Schwarz term  has a positive effect
on the gaugino masses, however the scalar masses are still dominated by $V_0$
and  the problem of tachyons persists.

Assuming that the above problems are solved one way or another, we can study,
at least qualitatively, the CP phases in the model. For the racetrack model we have
\begin{eqnarray}
{\rm Arg}(M_a) &=& 2.147\;, \nonumber\\
{\rm Arg}(A_{\alpha\beta\gamma}) &=& -1.387\;, \nonumber\\
{\rm Arg}(\hat\mu) &=& -0.041\;, \nonumber\\
{\rm Arg}(B \hat\mu) &=& 0\;.
\label{ph}
\end{eqnarray}
For the S-dual model the CP phases are very similar.
The resulting physical phases are
\begin{eqnarray}
{\rm Arg}\biggl( (B\hat \mu)^*\hat \mu M \biggr) &\simeq &2.1  \;,\nonumber\\
{\rm Arg}\left( A^*M \right) &\simeq& 0.4 \;.
\label{phys}
\end{eqnarray}
We see that generically the induced phases are ${\cal O}(1)$  and we
encounter the SUSY CP problem which is the subject of our next section.

\section{Electric Dipole Moments}

In heterotic string models there are three types of contributions to the electric
dipole moments. The first of them is the standard contribution from
complex phases in $F_{S,T}$ and $\mu$. The second appears due to nonuniversality even
if  $F_{S,T}$ and $\mu$ are real. The last contribution is induced by Im$S$ which
generates the $\bar \theta_{QCD}$ term. Let us consider each of these contributions
in more detail.

{\bf i. Complex phases in $F_{S,T}$ and $\mu$.}

These are the well known contributions originating from complex phases in
the gaugino masses, A-terms, the $\mu$ and $B\mu$ terms. The electron, neutron,
and mercury EDMs impose the following constraints on these complex phases (at the GUT scale)
\cite{Abel:2001vy}:
\begin{eqnarray}
&& \phi_A \leq 10^{-2}-10^{-1} \;, \nonumber\\
&& \phi_{\mu} \leq 10^{-3}-10^{-2} \;, \nonumber\\
&& \phi_{gaug.} \leq 10^{-2} \;.
\end{eqnarray}
Here $m_{3/2}$ is assumed to be of order $200$ GeV. To obtain each of these bounds
all the phases except for the one under consideration  have been set to zero.
Clearly, the complex phases in Eqs.\ref{ph} and \ref{phys} violate these bounds
and induce large EDMs unless the soft masses are pushed up to 10 TeV.

{\bf ii. Nonuniversality.}

In string models the SUSY CP problem appears to be more severe than in general
supersymmetric models. The reason is that, if no spontaneously broken supergravity is assumed,
the A-terms and the Yukawa matrices do not have to be related.
One can treat these on different grounds and entirely separate the Standard Model from
the rest of the MSSM.
This is not the case in string models. Specifically, the A-terms have a contribution
from the Yukawa couplings which is proportional to  $\partial_m \ln Y_{\alpha \beta \gamma}$.
Since the Yukawa matrices have a complicated flavour structure, the same is true
for the A-terms. This is indeed what happens in our example: if we are to reproduce
 CP violation in the Standard Model, we are bound to place the quark fields at different
orbifold fixed points. The corresponding Yukawa couplings are necessarily T-dependent and
 the A-terms   non-universal. As we will show, this leads to
unacceptably  large electric dipole moments even if  $F_{S,T}$ and the soft terms are
completely $real$.

Let us first note that the relevant quantities appearing in the soft Lagrangian are
\begin{equation}
  \hat A_{\alpha \beta \gamma}= A_{\alpha \beta \gamma}  Y_{\alpha \beta \gamma}\;.
\label{a}
\end{equation}
Clearly, these quantities are $necessarily$ complex due to the
complex phases in the Yukawa matrices. This would not be dangerous
for the EDMs were the A-terms universal and real. What matters is
the complex phases in the squark mass  insertions in the super-CKM
basis, i.e. in the basis  where the Yukawa matrices are diagonal.
To draw a correspondence between the supergravity notation we have
used above and the ``phenomenological'' notation, let us fix the
first index of the Yukawa to refer to the Higgs fields, the second
index to be the generational index for the left-handed fields, and
the last index to be that for  the right-handed fields. For
example, $Y_{{H_1}Q_iD_j}\equiv Y^d_{ij}$. Then, the super-CKM
basis is defined by
\begin{eqnarray}
&& \hat U_{L,R} \rightarrow V^u_{L,R}~ \hat U_{L,R} \;\;,\;\;
   \hat D_{L,R} \rightarrow V^d_{L,R}~ \hat D_{L,R} \;,\;\nonumber\\
&&  Y^u \rightarrow V_L^{u~T}~ Y^u~ V_R^{u*}={\rm diag}(h_u,h_c,h_t)\;,\;\nonumber\\
&&  Y^d \rightarrow V_L^{d~T}~ Y^d~ V_R^{d*}={\rm diag}(h_d,h_s,h_b)\;,
\label{super}
\end{eqnarray}
where $\hat U,\hat D$ are the quark superfields.
The matrices  $\hat A^{u,d}$ from Eq.\ref{a} are not diagonal in this basis and
their diagonal elements generically contain order one complex phases.
This is to be contrasted with the universal case where $\hat A^{u,d}$
and $Y^{u,d}$ are diagonal simultaneously with the former being real if
$F_{S,T}$ are real.

Complex phases in the diagonal elements of $\hat A^{u,d}$ in the super CKM basis 
induce electric dipole moments of the quarks.
In the universal case, the diagonal entries are proportional to the corresponding quark
masses. For instance,  $\hat A^u_{11}$ is much smaller than $\hat A^u_{22}$, etc.
In this case, the complex phases are required to be less than $10^{-1}-10^{-2}$
\cite{Abel:2001vy}.
The constraints become much stronger in the non-universal case. Indeed,
the diagonal entries of $\hat A^{u,d}$ in the super-CKM basis  are now proportional
to some linear combination of the quark masses. For instance,
\begin{equation}
\hat A_{11}^u \propto m_u + \epsilon m_c + \epsilon' m_t \;.
\end{equation}
This significantly increases the magnitude of $\hat A_{11}^{u,d}$.
Recall now that what matters for the EDMs is not just the phase of
 $\hat A_{11}^{u,d}$ but its imaginary part. Clearly, even a small
phase can be dangerous if the magnitude of $\hat A_{11}^{u,d}$ is
large.

In a non-universal case,
the EDM constraints can be  neatly expressed as constraints on the
imaginary parts of the so called  ``squark mass insertions''. These
are defined as $(\delta^{u,d}_{LR})_{ii}\sim \hat A^{u,d}_{ii} \langle H_{u,d} \rangle
/\tilde m^2$, where $\tilde m$ is the average squark mass
and the super-CKM basis is assumed.
The neutron EDM constrains their imaginary parts to be no greater than ${\cal O}(10^{-6})$,
whereas the mercury EDM constrains them to be less than ${\cal O}(10^{-7})$
\cite{Abel:2001vy}.
These bounds are violated in our case by orders of magnitude.
Indeed, with ${\cal O}(1)$ phases in the Yukawas and assuming ${\cal O}(10^{-2})$
mixing between the first and the third generations (i.e. $(V_{L,R})_{13}\sim V^{CKM}_{13}$),
we typically get Im$(\delta^{u,d}_{LR})_{11}\sim 10^{-4}$.
 Thus, the EDMs are
overproduced by two-three orders of magnitude.
Clearly, a similar effect occurs in the lepton sector if we allow non-diagonal lepton Yukawas.

This problem is quite generic for heterotic string models. If we are to produce
the CKM phase and fermion mass hierarchy, the Yukawas are bound to be nonuniversal
and T-dependent\footnote{Corrections from non-renormalizable operators do not change this.}.
This results in nonuniversal A-terms (unless $F_T=0$ which is highly
disfavoured by the dilaton stabilization mechanisms) inducing large EDMs
even in the absence of complex phases in $F_S$ and $F_T$.

{\bf iii. QCD vacuum angle $\bar \theta$.}

In addition to the above  sources of EDMs, there is a standard contribution to the $\bar \theta$
parameter from Im$S$. In the case of multiple gaugino condensates Im$S$ is fixed
(up to a discrete shift) by the potential minimization at an ${\cal O}(1)$ value.
So, generically this would overproduce the EDMs by many orders of magnitude.
To rectify this problem one needs a stringy Peccei-Quinn mechanism which would
set   $\bar \theta$ to zero regardless of its initial value. This requires
an anomalous global U(1) symmetry which couples to the QCD anomaly but not to those
of the  other condensing groups.  In string theory, such  symmetries  can arise from
anomalous gauge U(1)'s and reasonable solutions can be obtained \cite{Kim:1988dd}.
Here we will not address this issue in detail and will simply assume that the
strong CP problem is solved one way or another.

\section{Conclusions}

In this paper we have addressed the question whether it is  possible
to have spontaneous CP violation in heterotic string models at a
phenomenologically acceptable level. In addition to having CP violation,
we imposed the conditions of  dilaton stabilization and a viable
SUSY breaking scale.
We find the following positive features in the models considered:

{\bf +1.} CP can be broken by a VEV of the modulus field, while having
phenomenologically acceptable values for the dilaton  and the SUSY
breaking scale.

{\bf +2.} A non-trivial CKM phase can be produced.

Despite these encouraging general results, we encounter a number
of difficulties to be solved in more realistic models:

{\bf -1.} ${\cal O}(1)$ complex phases in the Yukawa matrices and $F_T$
lead to the EDMs exceeding the experimental limits by orders
of magnitude.

{\bf -2.} $T$ is stabilized close to the fixed points of the modular group
which results in a suppressed CKM phase.

{\bf -3.} Generally there are  tachyons and unacceptably light gauginos
(the latter problem is mitigated in the presence of the Green-Schwarz term).

{\bf -4.} Im$S$ gives a large contribution to $\bar \theta$ leading to the strong
CP problem (which can be resolved in the presence of a global anomalous $U(1)$
symmetry).

In addition, it is very difficult to obtain a vanishing cosmological constant while
retaining the positive features of the model. This however may not be a real problem
since $V_0$ is not necessarily directly related to the cosmological constant.

Some of these problems can be solved if supersymmetry breaking is dilaton-dominated,
i.e. $F_S \not=0$ and $F_T \simeq 0$. This would suppress the EDM contributions
originating from the non-universality and also help avoid tachyons and light
gauginos. At the same time, in order to produce the CKM phase, the modulus field must
be well away from the fixed points. Even if all this is the case, one has to ensure
that there are  no relative phases  among $\mu$, $B\mu$, and the gaugino masses  which is quite
nontrivial even in the dilaton-dominated case. We find that this is hardly possible
with the presently available dilaton stabilization mechanisms. In principle one can
combine different mechanisms to obtain the desired features. We will report
on this in a subsequent paper.

{\bf Acknowledgements.} The authors are indebted to David Bailin and George 
Kraniotis  for numerous valuable discussions. We have also benefited from communications
with Tom Dent and Joel Giedt.

\newpage

\begin{table}
\begin{center}
\vspace{0.5cm}
\begin{tabular}{|c|c||c|c|c|c||}\hline
$m$& $n$ &$S_{min}$  &$T_{min}$&$F_S$&$F_T$ \\ \hline\hline
0&0&$2.1299-0.9196i$&1.2346&$0$&$2.16 \times 10^{-16}$\\
0&1&$2.1299-0.9196i$&1.0000&$0$&$0$\\

0&2&$2.1299-0.9196i$&1.0000&$0$&$0$\\

0&3&$2.1299-0.9196i$&1.0000&$0$&$0$\\

0&4&$2.1299-0.9196i$&1.0000&$0$&$0$\\

0&5&$2.1299-0.9196i$&1.0000&$0$&$0$\\

1&0&$2.1299-0.9196i$&$0.9850~e^{0.5417i}$&0&$(-0.51-2.74i)\times 10^{-16}$\\

1&1&$2.1299-0.9196i$&$1.0000~e^{0.2401i}$&$0$&$0$\\

1&2&$2.1299-0.9196i$&$1.0000~e^{0.1913i}$&$0$&$0$\\

1&3&$2.1299-0.9196i$&$1.0000~e^{0.1642i}$&$0$&$0$\\

1&4&$2.1299-0.9196i$&$1.0000~e^{0.1462i}$&$0$&$0$\\

1&5&$2.1299-0.9196i$&$1.0000~e^{0.1331i}$&$0$&$0$\\

2&0&$2.1299-0.9196i$&$0.9922~e^{0.5329i}$&0&$(-1.08-5.94i)\times 10^{-15}$\\

2&1&$2.1299-0.9196i$&$1.0000~e^{0.2897i}$&$0$&$0$\\

2&2&$2.1299-0.9196i$&$1.0000~e^{0.2412i}$&$0$&$0$\\

2&3&$2.1299-0.9196i$&$1.0000~e^{0.2121i}$&$0$&$0$\\

2&4&$2.1299-0.9196i$&$1.0000~e^{0.1919i}$&$0$&$0$\\

2&5&$2.1299-0.9196i$&$1.0000~e^{0.1766i}$&$0$&$0$\\

3&0&$2.1299-0.9196i$&$0.9972~e^{0.5159i}$&$0$&$(-1.32-1.09i)\times 10^{-13}$\\

3&1&$2.1299-0.9196i$&$1.0000~e^{0.3168i}$&$0$&$0$\\

3&2&$2.1299-0.9196i$&$1.0000~e^{0.2705i}$&$0$&$0$\\

3&3&$2.1299-0.9196i$&$1.0000~e^{0.2416i}$&$0$&$0$\\

3&4&$2.1299-0.9196i$&$1.0000~e^{0.2209i}$&$0$&$0$\\

3&5&$2.1299-0.9196i$&$1.0000~e^{0.2049i}$&$0$&$0$\\

4&0&$2.1299-0.9196i$&$0.9960~e^{0.5283i}$&$0$&$(-0.94-5.24i)\times 10^{-12}$\\

4&1&$2.1299-0.9196i$&$1.0000~e^{0.3347i}$&0&0\\

4&2&$2.1299-0.9196i$&$1.0000~e^{0.2906i}$&$0$&$0$\\

4&3&$2.1299-0.9196i$&$1.0000~e^{0.2624i}$&$0$&$0$\\

4&4&$2.1299-0.9196i$&$1.0000~e^{0.2418i}$&$0$&$0$\\

4&5&$2.1299-0.9196i$&$1.0000~e^{0.2257i}$&$0$&$0$\\

5&0&$2.1299-0.9196i$&$0.9968~e^{0.5274i}$&$0$&$(-0.31-1.76i)\times 10^{-10}$\\

5&1&$2.1299-0.9196i$&$1.0000~e^{0.3478i}$&0&0\\

5&2&$2.1299-0.9196i$&$1.0000~e^{0.3056i}$&0&0\\

5&3&$2.1299-0.9196i$&$1.0000~e^{0.2782i}$&$0$&$0$\\

5&4&$2.1299-0.9196i$&$1.0000~e^{0.2579i}$&$0$&$0$\\

5&5&$2.1299-0.9196i$&$1.0000~e^{0.2419i}$&$0$&$0$\\

\hline
\end{tabular}
\caption{Minima for the racetrack models, $\delta_{GS}=0$.}\label{table1}
\end{center}
\end{table}


\newpage

\begin{table}[h]
\begin{center}
\vspace{0.5cm}
\begin{tabular}{|c|c||c|c|c|c||}\hline
$m$& $n$ &$S_{min}$  &$T_{min}$&$F_S$&$F_T$ \\ \hline\hline
0&0&$1.8843-0.9196i$&$1.2326$&$-1.43\times 10^{-17}$&$1.76\times 10^{-17}$\\
0&1&$1.9295-0.9196i$&1.0000&$0$&$0$\\

0&2&$1.9295-0.9196i$&1.0000&$0$&$0$\\

0&3&$1.9295-0.9196i$&1.0000&$0$&$0$\\

0&4&$1.9295-0.9196i$&1.0000&$0$&$0$\\

0&5&$1.9295-0.9196i$&1.0000&$0$&$0$\\

1&0&$1.9608-1.0137i$&$0.9918~e^{0.5005i}$&$(1.65-1.33i)\times 10^{-17}$&$(-1.85+1.45i)\times 10^{-17}$\\

1&1&$1.9365-0.9652i$&$1.0000~e^{0.2400i}$&0&0\\

1&2&$1.9340-0.9559i$&$1.0000~e^{0.1912i}$&0&0\\

1&3&$1.9328-0.9507i$&$1.0000~e^{0.1641i}$&0&0\\

1&4&$1.9321-0.9473i$&$1.0000~e^{0.1462i}$&0&0\\

1&5&$1.9317-0.9448i$&$1.0000~e^{0.1331i}$&0&0\\

2 &   0 &$ 1.9604 -1.0163 i$&$   0.9958~e^{ 0.5117 i}$&$(3.55-2.92i)\times 10^{-16}$&$(-3.95+3.20i)\times 10^{-16}$ \\ 

 2 &   1 &$ 1.9396 -0.9746 i$&$   1.0000 ~e^{ 0.2897 i}$&0&0 \\ 

2 &   2 &$ 1.9366 -0.9654 i$&$   1.0000 ~e^{ 0.2412 i}$&0&0 \\ 

2 &   3 &$ 1.9350 -0.9599 i$&$   1.0000 ~e^{ 0.2121 i}$&0&0 \\ 

2 &   4 &$ 1.9340 -0.9560 i$&$   1.0000 ~e^{ 0.1919 i}$&0&0 \\ 

 2 &   5 &$ 1.9333 -0.9531 i$&$   1.0000 ~e^{ 0.1766 i}$&0&0 \\

3 &   0 &$ 1.9600 -1.0174 i$&$   0.9998 ~e^{ 0.5151 i}$&$(6.85-4.04i)\times 10^{-16}$&$(-7.64+4.42i)\times 10^{-16}$ \\ 

3 &   1 &$ 1.9416 -0.9797 i$&$   1.0000 ~e^{ 0.3167 i}$&0&0 \\ 

3 &   2 &$ 1.9384 -0.9709 i$&$   1.0000 ~e^{ 0.2704 i}$&0&0 \\

3 &   3 &$ 1.9366 -0.9655 i$&$   1.0000 ~e^{ 0.2416 i}$&0&0 \\

 3 &   4 &$ 1.9355 -0.9615 i$&$   1.0000 ~e^{ 0.2208 i}$&0&0 \\

 3 &   5 &$ 1.9346 -0.9585 i$&$   1.0000 ~e^{ 0.2049 i}$&0&0 \\

 4 &   0 &$ 1.9608 -1.0176 i$&$  0.9969 ~e^{ 0.5180 i}$&$(4.03-3.93i)\times 10^{-13}$&$(-4.47+4.32i)\times 10^{-13}$ \\

4 &   1 &$ 1.9429 -0.9831 i$&$   1.0000 ~e^{ 0.3347 i}$&0&0 \\ 

 4 &   2 &$ 1.9397 -0.9748 i$&$   1.0000 ~e^{ 0.2906 i}$&0&0 \\

  4 &   3 &$ 1.9379 -0.9694 i$&$   1.0000 ~e^{ 0.2624 i}$&0&0 \\ 

 4 &   4 &$ 1.9366 -0.9655 i$&$   1.0000 ~e^{ 0.2417 i}$&0&0 \\

4 &   5 &$ 1.9357 -0.9624 i$&$   1.0000 ~e^{ 0.2256 i}$&0&0 \\   

5 &   0 &$ 1.9613 -1.0180 i$&$   0.9959 ~e^{ 0.5205 i}$&$(1.49-2.24i)\times 10^{-11}$&$(-1.65+2.47i)\times 10^{-11}$ \\ 

 5 &   1 &$ 1.9440 -0.9856 i$&$   1.0000 ~e^{ 0.3477 i}$&0&0 \\ 

 5 &   2 &$ 1.9408 -0.9776 i$&$   1.0000 ~e^{ 0.3056 i}$&0&0 \\ 

 5 &   3 &$ 1.9389 -0.9724 i$&$   1.0000 ~e^{ 0.2782 i}$&0&0 \\ 

 5 &   4 &$ 1.9376 -0.9686 i$&$   1.0000 ~e^{ 0.2579 i}$&0&0 \\ 

5 &   5 &$ 1.9366 -0.9655 i$&$   1.0000 ~e^{ 0.2419 i}$&0&0 \\

\hline
\end{tabular}\\
\caption{Minima for the racetrack models, $\delta_{GS}=5$.}\label{table2}
\end{center}

\end{table}

\newpage

\begin{table}[h]
\begin{center}
\vspace{0.5cm}
\begin{tabular}{|c|c||c|c|c|c||}\hline
$m$& $n$ &$S_{min}$  &$T_{min}$& $F_S$ &$F_T$\\ \hline\hline
0&0&1.8&$~e^{i \pi/6}$&$-5.8298\times 10^{-7}$&0\\
0&1&1.8&$~e^{i \pi/6}$&0&0\\
0&2&1.8&$~e^{i \pi/6}$&0&0\\
0&3&1.8&$~e^{i \pi/6}$&0&0\\
0&4&1.8&$~e^{i \pi/6}$&0&0\\
0&5&1.8&$~e^{i \pi/6}$&0&0\\
1&0&1.8&1&0&0\\

1&1&1.8&$1,~e^{i \pi/6}$&0&0\\

1&2&1.8&$1,~e^{i \pi/6}$&0&0\\

1&3&1.8&$1,~e^{i \pi/6}$&0&0\\

1&4&1.8&$1,~e^{i \pi/6}$&0&0\\

1&5&1.8&$1,~e^{i \pi/6}$&0&0\\
2&0&1.8&1&0&0\\

2&1&1.8&$1,~e^{i \pi/6}$&0&0\\

2&2&1.8&$1,~e^{i \pi/6}$&0&0\\

2&3&1.8&$1,~e^{i \pi/6}$&0&0\\

2&4&1.8&$1,~e^{i \pi/6}$&0&0\\

2&5&1.8&$1,~e^{i \pi/6}$&0&0\\
3&0&$1.8$&1&0&0\\

3&1&1.8&$1,~e^{i \pi/6}$&0&0\\

3&2&1.8&$1,~e^{i \pi/6}$&0&0\\                              

3&3&1.8&$1,~e^{i \pi/6}$&0&0\\

3&4&1.8&$1,~e^{i \pi/6}$&0&0\\

3&5&1.8&$1,~e^{i \pi/6}$&0&0\\

4&0&$1.8$&1&0&0\\

4&1&1.8&$1,~e^{i \pi/6}$&0&0\\

4&2&1.8&$1,~e^{i \pi/6}$&0&0\\

4&3&1.8&$1,~e^{i \pi/6}$&0&0\\

4&4&1.8&$1,~e^{i \pi/6}$&0&0\\

4&5&1.8&$1,~e^{i \pi/6}$&0&0\\
5&0&1.8&1&0&0\\

5&1&1.8&$1,~e^{i \pi/6}$&0&0\\
5&2&1.8&$1,~e^{i \pi/6}$&0&0\\

5&3&1.8&$1,~e^{i \pi/6}$&0&0\\

5&4&1.8&$1,~e^{i \pi/6}$&0&0\\

5&5&1.8&$1,~e^{i \pi/6}$&0&0\\

\hline
\end{tabular}\\

\caption{ Minima for models with the non-perturbative K\"{a}hler potential,
$\delta_{GS}=0$.}\label{table3}
\end{center}

\end{table}

\newpage
\begin{table}[h]
\begin{center}
\vspace{0.5cm}
\begin{tabular}{|c|c||c|c|c|c||}\hline
$m$& $n$ &$S_{min}$  &$T_{min}$& $F_S$ &$F_T$\\ \hline\hline
0&0&$--$&$0$&$--$&$--$\\
0&1&1.9&$~e^{i \pi/6}$&0&0\\
0&2&1.9&$~e^{i \pi/6}$&0&0\\
0&3&1.9&$~e^{i \pi/6}$&0&0\\
0&4&1.9&$~e^{i \pi/6}$&0&0\\
0&5&1.9&$~e^{i \pi/6}$&0&0\\
1&0&1.8&1&0&0\\

1&1&1.8, 1.9&$1,~e^{i \pi/6}$&0&0\\

1&2&1.8, 1.9&$1,~e^{i \pi/6}$&0&0\\

1&3&1.8, 1.9&$1,~e^{i \pi/6}$&0&0\\

1&4&1.8, 1.9&$1,~e^{i \pi/6}$&0&0\\

1&5&1.8, 1.9&$1,~e^{i \pi/6}$&0&0\\
2&0&1.8&1&0&0\\

2&1&1.8, 1.9&$1,~e^{i \pi/6}$&0&0\\

2&2&1.8, 1.9&$1,~e^{i \pi/6}$&0&0\\

2&3&1.8, 1.9&$1,~e^{i \pi/6}$&0&0\\

2&4&1.8, 1.9&$1,~e^{i \pi/6}$&0&0\\

2&5&1.8, 1.9&$1,~e^{i \pi/6}$&0&0\\
3&0&$1.8$&1&0&0\\

3&1&1.8, 1.9&$1,~e^{i \pi/6}$&0&0\\

3&2&1.8, 1.9&$1,~e^{i \pi/6}$&0&0\\

3&3&1.8, 1.9&$1,~e^{i \pi/6}$&0&0\\
3&4&1.8, 1.9&$1,~e^{i \pi/6}$&0&0\\

3&5&1.8, 1.9&$1,~e^{i \pi/6}$&0&0\\

4&0&$1.8$&1&0&0\\

4&1&1.8, 1.9&$1,~e^{i \pi/6}$&0&0\\

4&2&1.8, 1.9&$1,~e^{i \pi/6}$&0&0\\

4&3&1.8, 1.9&$1,~e^{i \pi/6}$&0&0\\

4&4&1.8, 1.9&$1,~e^{i \pi/6}$&0&0\\

4&5&1.8, 1.9&$1,~e^{i \pi/6}$&0&0\\
5&0&1.8&1&0&0\\

5&1&1.8, 1.9&$1,~e^{i \pi/6}$&0&0\\

5&2&1.8, 1.9&$1,~e^{i \pi/6}$&0&0\\

5&3&1.8, 1.9&$1,~e^{i \pi/6}$&0&0\\

5&4&1.8, 1.9&$1,~e^{i \pi/6}$&0&0\\

5&5&1.8, 1.9&$1,~e^{i \pi/6}$&0&0\\

\hline
\end{tabular}\\

\caption{ Minima for models with the non-perturbative K\"{a}hler potential,
$\delta_{GS}=5$. For the parameter values considered, in the $m=n=0$ case
the extra dimensions become uncompactified.  }\label{table4}
\end{center}

\end{table}

\newpage

\begin{table}
\begin{center}
\vspace{0.5cm}
\begin{tabular}{|c|c||c|c|c|c||}\hline
$m$& $n$ &$S_{min}$  &$T_{min}$&$F_S$&$F_T$ \\ \hline\hline
0&0&$1$&1.2346&$0$&$4.24 \times 10^{-16}$\\
0&1&$1$&1.0000&$0$&$0$\\

0&2&$1$&1.0000&$0$&$0$\\

0&3&$1$&1.0000&$0$&$0$\\

0&4&$1$&1.0000&$0$&$0$\\

0&5&$1$&1.0000&$0$&$0$\\

1&0&$1$&$0.9850~e^{0.5417i}$&0&$(-1.00-5.38i)\times 10^{-16}$\\

1&1&$1$&$1.0000~e^{0.2401i}$&$0$&$0$\\

1&2&$1$&$1.0000~e^{0.1913i}$&$0$&$0$\\

1&3&$1$&$1.0000~e^{0.1642i}$&$0$&$0$\\

1&4&$1$&$1.0000~e^{0.1462i}$&$0$&$0$\\

1&5&$1$&$1.0000~e^{0.1331i}$&$0$&$0$\\

2&0&$1$&$0.9922~e^{0.5329i}$&0&$(-0.21-1.17i)\times 10^{-14}$\\

2&1&$1$&$1.0000~e^{0.2897i}$&$0$&$0$\\

2&2&$1$&$1.0000~e^{0.2412i}$&$0$&$0$\\

2&3&$1$&$1.0000~e^{0.2121i}$&$0$&$0$\\

2&4&$1$&$1.0000~e^{0.1919i}$&$0$&$0$\\

2&5&$1$&$1.0000~e^{0.1766i}$&$0$&$0$\\

3&0&$1$&$0.9972~e^{0.5159i}$&$0$&$(-2.60-2.14i)\times 10^{-13}$\\
3&1&$1$&$1.0000~e^{0.3168i}$&$0$&$0$\\

3&2&$1$&$1.0000~e^{0.2705i}$&$0$&$0$\\

3&3&$1$&$1.0000~e^{0.2416i}$&$0$&$0$\\

3&4&$1$&$1.0000~e^{0.2209i}$&$0$&$0$\\

3&5&$1$&$1.0000~e^{0.2049i}$&$0$&$0$\\

4&0&$1$&$0.9960~e^{0.5283i}$&$0$&$(-0.18-1.03i)\times 10^{-11}$\\

4&1&$1$&$1.0000~e^{0.3347i}$&0&0\\

4&2&$1$&$1.0000~e^{0.2906i}$&$0$&$0$\\

4&3&$1$&$1.0000~e^{0.2624i}$&$0$&$0$\\

4&4&$1$&$1.0000~e^{0.2418i}$&$0$&$0$\\

4&5&$1$&$1.0000~e^{0.2257i}$&$0$&$0$\\

5&0&$1$&$0.9968~e^{0.5274i}$&$0$&$(-0.62-3.45i)\times 10^{-10}$\\

5&1&$1$&$1.0000~e^{0.3478i}$&0&0\\

5&2&$1$&$1.0000~e^{0.3056i}$&0&0\\

5&3&$1$&$1.0000~e^{0.2782i}$&$0$&$0$\\

5&4&$1$&$1.0000~e^{0.2579i}$&$0$&$0$\\

5&5&$1$&$1.0000~e^{0.2419i}$&$0$&$0$\\

\hline
\end{tabular}
\caption{Minima for S-dual models, $\mu=1.8\times 10^{-3}$.}\label{table5}
\end{center}
\end{table}


\end{document}